\documentclass{elsart1p}
\usepackage{graphicx}
\usepackage{amssymb}
\begin{document}
\begin{frontmatter}
%
%
%
\title{Lattice QCD}
%
%
\author{Sinya Aoki}
\address{Graduate School of Pure and Applied Sciences, University of Tsukuba, Tsukuba, Ibaraki 305-8571, Japan\\
Riken-BNL Research Center, BNL, Upton, NY11973, USA}
\begin{abstract}
Some topics from recent progresses in lattice QCD are reviewed. 
\end{abstract}
\end{frontmatter}
%
\section{Lattice QCD at $T=0$} 
\subsection{Full QCD simulations near or at the physical quark mass}
Due to large computational costs, $u$, $d$ quark mass in current lattice QCD simulations is heavier than the physical value.
One then has to extrapolate results obtained at heavier quark mass to those at the physical point, using some formula such as chiral perturbation theory(ChPT). It is not clear, however, whether ChPT can be applied at rather heavy quark masses employed so far in lattice QCD simulations.   
This situation is gradually changing,  thanks to improvements for both computers and algorithms.
Indeed the first important message from the annual lattice conference of this year, lattice 2008, is that full QCD simulations near the physical light quark mass becomes possible.

In table\ref{tab:full_QCD}, parameters such as number of flavors, the lattice spacing $a$ , the spatial extension of the lattice $L$, the minimum pion mass $m_\pi^{\rm min.}$ and $ L\times m_\pi^{\rm min.}$, are listed for recent large scale full QCD simulations.
In the left half of the list, the conventional quark action such as the Wilson quark action or the staggered quark action is employed for simulations, while the chirally symmetric quark action such as the overlap quark action or the domain-wall quark action is used in the right half. Let me point out here that the minimum pion mass in the full QCD simulations now becomes below 200 MeV for conventional actions and close to 300 MeV for chirally symmetric actions, though finite size effects could be sizable in some cases where $L m_\pi < 3$.
\begin{table}[bt]
\label{tab:full_QCD}
\caption{(Incomplete) list of recent full QCD simulations.}
\begin{tabular}{|c|c|c|c|c||c| c|c|c|c|}
\hline
\hline
\multicolumn{5}{| l ||}{Conventional quark action} &
\multicolumn{5}{| l |}{Chirally symmetric quark action}\\
\hline
Group & $a$(fm) & $L$(fm) & $m_\pi^{\rm min}$(MeV) & $ L m_\pi^{\rm min}$ &
Group & $a$(fm) & $L$(fm) & $m_\pi^{\rm min}$(MeV) & $ L m_\pi^{\rm min}$ \\
\hline
2+1 flavors & & & & 
&2+1 flavors & & & &  \\
\hline
PACS-CS\cite{pacs-cs}  & 0.09 & 2.9 & 160 &2.3 & RBC-UKQCD\cite{rbc-ukqcd}   & 0.11 & 2.8 & 330 & 4.6\\
MILC\cite{milc}  & $\ge 0.06$ & 3.3 & 240 & 4 &JLQCD\cite{jlqcd_nf3}   & 0.11 & 1.8 & 315 & 2.8\\
BMW\cite{bmw}  & $\ge 0.065$ & $ \ge 4.2$ & 190 & 4 & & & & &  \\
\hline
2 flavors & & & & &
2 flavors & & & & \\
\hline
CERN-ToV\cite{cern}&   $\ge 0.05$ & 1.7-1.9 & 300& 2.9 & RBC\cite{rbc}  & 0.12 & 2.5 & 490 & 6.1\\
ETMC\cite{etmc} & $\ge0.07$ & 2.1 & 300 &  3.2  & JLQCD\cite{jlqcd_nf2}  & 0.12 & 1.9 & 290& 2.8\\
CLS\cite{cls} & $ 0.08$ & 2.6 & 230 & 3 & & & & &   \\
QCDSF\cite{qcdsf} & $\ge 0.072$ & $2.3$ & 240 & 2.8   & & & & &  \\  
\hline
\hline
\end{tabular}
\end{table}

\begin{figure}[b]
\centering
\includegraphics[width=55mm,angle=270,clip]{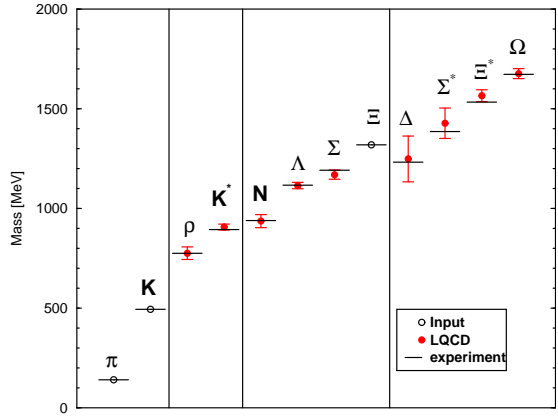}
\caption{Hadron spectrum in the continuum limit. $m_\pi$, $m_K$ and $m_\Xi$ are used as inputs 
to fix free parameters of QCD.}
\label{fig:spectrum}
\end{figure}

Results for hadron spectrum from one of the most extensive calculations have been reported in Ref.\cite{bmw}, where both chiral and continuum extrapolations have been performed. 
In this calculation, the minimum pion mass $m_\pi^{\rm min.} = 190$ MeV is light enough, and $m_\pi L \ge 4$ is always satisfied.  In Fig.\ref{fig:spectrum}, the hadron spectrum obtained in the continuum limit is compared with experiment. Here the overall scale, the light quark mass and the strange quark mass are fixed by $m_\Xi$, $m_\pi$ and $m_K$. As can be seen from the figure, the agreement between lattice QCD and experiment is excellent.

\begin{figure}[bt]
\centering
\includegraphics[width=55mm,angle=270,clip]{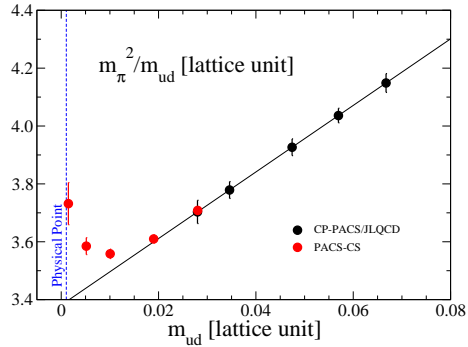}
\caption{$m_\pi^2/m_{ud}$ as a function of $m_{ud}$ at $a=0.09$ fm and $L=2.9$ fm. 
Red circles correspond to $m_\pi = 702$, 570, 412, 296, 156 MeV from right to left. The vertical dashed line is the physical point. }
\label{fig:pacs-cs}
\end{figure}

It is also reported recently that the pion mass can be smaller than 160 MeV in the 2+1 flavor full QCD simulation at $a=0.09$ fm\cite{pacs-cs}. 
In Fig. \ref{fig:pacs-cs}, the pion mass squired divided by the light quark mass, $m_\pi^2/m_{ud}$, is plotted as a function of the light quark mass, where black solid circles are previous results from CP-PACS/JLQCD collaborations\cite{cppacs-jlqcd} obtained at heavier quark masses while red solid circles are newt ones from PACS-CS collaboration\cite{pacs-cs}.
As quark mass decreases, this ratio increasingly deviates from the straight line obtained from the fit with black circles, suggesting an existence of the chiral logarithm in the ratio.  Note however that
$L m_\pi^{\rm min.} = 2.3$ in this calculation, so that the finite size effect might be sizable at the lightest quark mass. To reduce the magnitude of the finite size effect less than a \% level, gauge configurations at $L=5.8$ fm with $m_\pi \simeq 140$ MeV corresponding $m_\pi L \simeq 4.1$ are currently accumulated by this group. This simulation indeed will become the real QCD calculation.

\subsection{Lattice QCD and chiral perturbation theories}
The second message of this year from lattice QCD is that NLO SU(2) chiral perturbation theory(ChPT) seems to work at $m_\pi\le 450$ MeV while NLO SU(3) ChPT seems to fail at the physical strange quark mass. The first positive statement can be established by both 2-flavor full QCD simulations with the small enough quark mass and use of  chirally symmetric quark such as overlap or domain-wall
quark. The second negative statement has been  concluded from analyses in 2+1 flavor full QCD simulations. As a consequence SU(2) ChPT analysis is applied even to data in 2+1 flavor QCD.

\begin{figure}[bt]
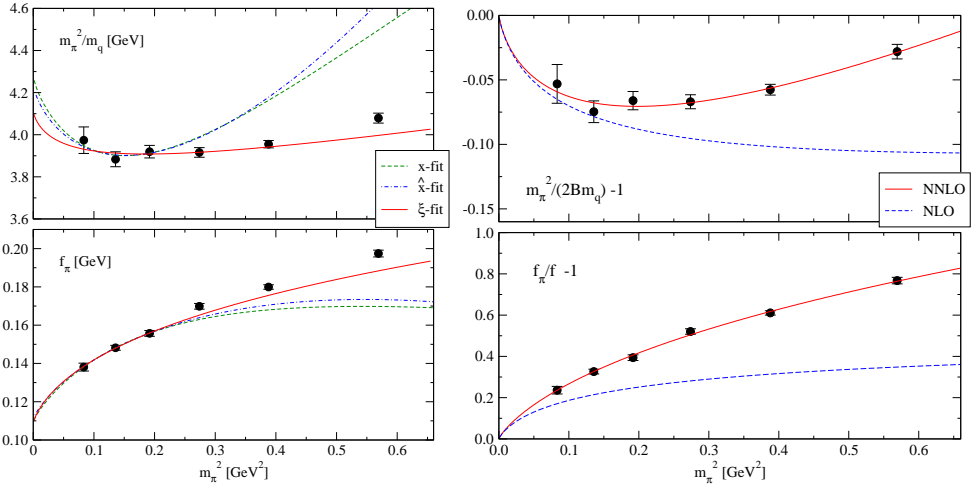

\centering
\includegraphics[width=60mm,clip]{Figs/JLQCD_mpfp_vp_nlo_3pts.eps}
\includegraphics[width=68mm,clip]{Figs/JLQCD_converge_6pts.eps}
\caption{Left: $m_\pi^2/m_q$(top) and $f_\pi$ (bottom) as a function of $m_\pi^2$ at $a=0.12$ fm. 
Right: NNLO chiral fit with $M=m_\pi^2/(4\pi f)^2$ using all data points for $m_\pi^2/2B m_q-1$(top) and $f_\pi/f-1$(bottom). Solid curves are the full NNLO fit, while dashed ones represent the NLO contributions. }
\label{fig:chpt}
\end{figure}

In Fig.\ref{fig:chpt}  $m_\pi^2/m_q$(top-left) and $f_\pi$(bottom-left) are plotted as function of the quark mass $m_\pi^2$ with the 2-flavor overlap fermion at $a=0.12$ fm\cite{jlqcd_nf2}. The lightest point corresponds to $m_\pi = 290$ MeV while the heaviest to 720 MeV.
The lightest 3 points at $m_\pi \le 450$ MeV are fitted by NLO SU(2) ChPT formula given by
\begin{eqnarray}
\frac{m_\pi^2}{m_q} &=& 2B\left[ 1 + \frac{M}{2} \ln M \right]+c_3 M, \qquad
f_\pi = f\left[1-M \ln M\right] + c_4 M ,
\label{eq:chpt}
\end{eqnarray}
where
$M = m_q/(4\pi f)^2$ (green dash-dotted line), $M= m_\pi^2/(4\pi f)^2$ (blue dashed line) or
$M = m_\pi^2/(4\pi f_\pi)^2$ (red solid line).  Differences among these three choices are higher order of ChPT in eq.(\ref{eq:chpt}).  The fact that all three fits work reasonably well at $m_\pi \le 450$ MeV
establishes the validity of  the NLO SU(2) ChPT at $m_\pi < 450$ MeV.
It is also shown in Ref.\cite{jlqcd_chpt} that all data points up to the one at $m_\pi=720$ MeV can be fitted by the NNLO formula of the ChPT with $M = m_\pi^2/(4\pi f)^2$, whose NLO fit well describes data even beyond the fitted range. The convergence behaviour of the chiral expansion is also plotted for $m_\pi^2/m_q$(top-right) and $f_\pi$(bottom-right) in Fig.\ref{fig:chpt}, which shows, for example, that the NLO contribution is 
a -10\%( +28\%) level for $m_\pi^2/m_q$ ($f_\pi$) at  $m_\pi \simeq 500$ MeV, while the NNLO correction gives about +3\%(18\%).  Although the chiral expansion is at least convergent for both quantities, the NNLO contribution for $f_\pi$ becomes significant already at  $m_\pi \simeq 500$ MeV, corresponding to the $K$ meson mass region for the 3 flavor case.

In the 2+1 flavor full QCD simulations with the domain-wall quark, it is reported\cite{rbc-ukqcd} that the NLO SU(3) (partially quenched) ChPT can not explain the quark mass dependence of  mass and decay constant in the pseudo-scalar meson sector around the physical strange quark mass region.
Instead the formula in the NLO SU(2) ChPT plus the heavy strange quark theory works better even for these data in the 2+1 flavor full QCD. This fact is consistent with the conclusion in the previous paragraph  that the NNLO correction in the SU(2) ChPT can not be neglected at $m_\pi \simeq 500$ MeV for these quantities.

\subsection{Topology}
The 3rd message of this year from lattice QCD is that it becomes difficult to change topological charge in full QCD simulations at the lighter quark mass and/or near the continuum limit .
One of possible solutions to this problem recently investigated is  QCD with fixed topological charge.
\cite{fixQ_1,fixQ_2}. For example, the two point function between flavor singlet  pseudo-scalar densities
$P(x)$ at fixed topological charge $Q$ behaves at large separation as
\begin{eqnarray}
\lim_{\vert x\vert \rightarrow\infty}
\langle m P(x) m P(0) \rangle_Q &=& -\frac{\chi_t}{V} + \frac{1}{V^2}\left( Q^2 -\frac{c_4}{2\chi_t}\right) + O\left(\frac{1}{V^3}\right),
\label{eq:chi_t}
\end{eqnarray}
where $V$ is the space-time volume, $m$ is the quark mass, $\langle {\cal O} \rangle_Q$ is the expectation values at fixed $Q$, 
$
\chi_t = \langle Q^2\rangle/V$ and $c_4 =\langle Q^4\rangle_c/V
$
are the topological susceptibility and the 4-th cumulant defined in the ordinary QCD at $\theta = 0$.
This formula tells us that one can extract  informations about QCD at $\theta = 0$ such as $\chi_t$ or $c_4$ from behaviours of correlation functions calculated at fixed $Q$.  
This strategy is suitable for lattice QCD since it is easier to fix the topological charge $Q$ than to change
it correctly during simulations, in particular,  at small quark mass and/or near the continuum limit.

\begin{figure}[bt]
\centering
\includegraphics[width=58mm,clip]{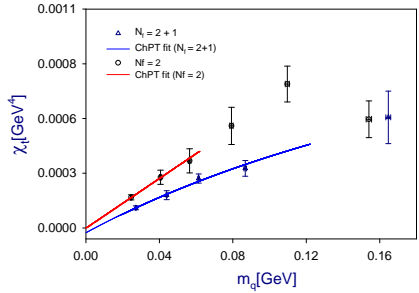}
\caption{ Topological susceptibility $\chi_t$ as a function of sea quark mass $m_q$ for $N_f=2$ (circles) and $N_f=2+1$ (triangles) at $Q=0$. The solid lines are fits with the form of ChPT.}
\label{fig:chi_t}
\end{figure}

The JLQCD has taken this strategy for the dynamical overlap project,  where the fixed topology not only guarantees the locality of the overlap Dirac operator but also makes full QCD simulations with overlap quarks  much faster. In Fig.\ref{fig:chi_t}, the topological susceptibility $\chi_t$ extracted from eq.(\ref{eq:chi_t}) at $Q=0$ is plotted as a function of the  light quark mass $m_q$ for 2 flavor\cite{chi_t_nf2} and 2+1 flavor\cite{chi_t_nf3} QCD. The results show a clear linear dependence on the sea quark mass
expected from the ChPT as $\chi_t = m_q\Sigma/2$ for 2 flavor and $\chi_t = m_q m_s \Sigma/(2m_s+ m_q)$ for 2+1 flavor, where $\Sigma$ is the chiral condensate at massless limit and $m_s$ is the (fixed) strange quark mass. The fits to data using this formula, denoted solid lines in the figure, yield,
after a renormalization,
$\Sigma({\rm 2\ GeV}) =[245(5)(10)\ {\rm MeV}]^3$ $(N_f=2)$ and
$\Sigma({\rm 2\ GeV}) =[240(5)(2)\ {\rm MeV}]^3$ $(N_f=3)$.

\section{Lattice QCD at finite temperature and density}
\subsection{Isentropic Equation of states at finite density}
It has been shown that the zero-viscosity (perfect) hydro calculations explain the RHIC experimental results well.  This indicates that in heavy-ion collision, after thermalization, the system expands and cools with constant entropy. In this case, the equation of state (EoS) along the line of the constant entropy per baryon number, called the isentropic EoS, becomes more relevant than the ordinary EoS as a function of temperature at the fixed chemical potential $\mu$. The entropy per baryon number, $S/N_B$ is roughly equal to 300 at RHIC, 45 at SPS and 30 at AGS.
The lattice QCD at non-zero chemical potential, however, is notoriously difficult to be simulated. So far there is no established method to simulate the full QCD at arbitrary value of the chemical potential $\mu$.  
If the chemical potential is small, the Taylor expansion in terms of  $\mu$ works well\cite{taylor}.

\begin{figure}[bt]
\centering
\includegraphics[width=58mm,clip]{Figs/MILC_Sisentr.eps}
\includegraphics[width=58mm,clip]{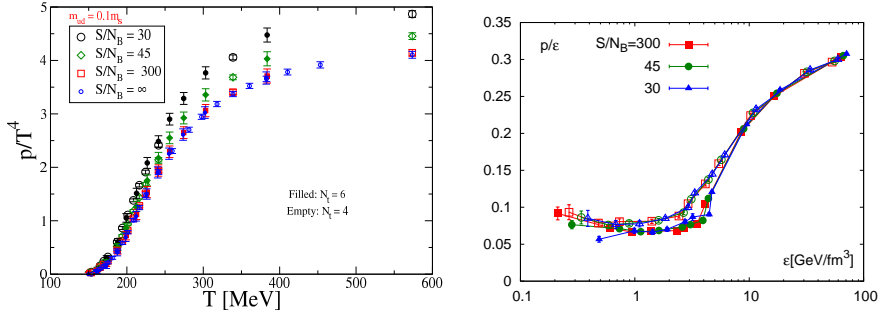}
\caption{ Left: Pressure $p/T^4$ as a function of  temperature $T$ along the lines of constant entropy per baryon number  in 2+1 flavor QCD\cite{milc2}. 
Right: The ratio of pressure and energy density $p/\epsilon$ as a function of $\epsilon$ along the lines of constant entropy per baryon number 2+1 flavor QCD\cite{rbc-b}.}
\label{fig:isentr}
\end{figure}

The MILC Collaboration has calculated the isentropic EoS using the Taylor expansion method in 2+1 flavor QCD with the staggered quarks at $m_\pi \simeq$ 220 MeV\cite{milc1,milc2}. In Fig.\ref{fig:isentr}(Left)
their result for $p/T^4$ where $p$ denotes the pressure is plotted as a function of $T$ along lines of the constant $S/N_B=$ 30, 45, 300 and $\infty$, at $N_t=4$ (open symbols) and $N_t=6$ (solid symbols) where
$N_t$ is the number of the lattice sites in the temporal  direction which controls the lattice spacing as
$a= 1/(N_t T)$.  The result shows that the lattice discretization errors are small for this quantity and
 $p/T^4$ becomes larger at fixed $T$ as $S/N_B$ decreases.

In Fig.\ref{fig:isentr}(Right) the result for $p/\epsilon$, obtained by RBC-Bielefeld Collaborations in 2+1 flavor QCD with the staggered quarks at $m_\pi \simeq$ 220 MeV using the Taylor expansion method\cite{rbc-b},
is plotted as a function of $\epsilon$ on lines of the constant $S/N_B =30, 45, 300$ at $N_t=4$ (filled) and $N_t=6$ (open), where $\epsilon$ is the energy density. Form this quantities the velocity of sound $c_s$, which is important for the hydrodynamics calculation in heavy ion collisions, can be extracted as
$
c_s^2 = \frac{d p}{d\epsilon} = \epsilon \frac{d(p/\epsilon)}{d \epsilon} + \frac{p}{\epsilon}
$ . 
Although discretization errors are observed, the dependence on $S/N_B$ seems small in this ratio.

\subsection{QCD critical point at finite density}
The finite temperature phase "transition" 
in 2+1 flavor QCD at the physical light (u,d) quark mass and the physical strange quark mass
seems a cross-over at zero or small chemical potential, while the phase transition is expected to become
of first order at large chemical potential. Between the cross-over region and the first order region, there must exist an unique second order phase transition point, which is called the critical end point of QCD.
Important questions to be answered are as follows. Does the critical end point indeed exist ? If so at which temperature and density does it occur ?
To answer these questions, lattice QCD simulations at large $\mu$ are required.
Unfortunately the Taylor expansion method mentioned before does not work at such large $\mu$.
The reweighting method, which looked promising on small volumes, may not be reliable to answer these questions\cite{ejiri1}.  A new method appears\cite{ejiri2}, but we have to wait for a while, in order to see whether the method indeed works for this problem.
One established method to investigate the existence of the critical end point is the imaginary chemical potential method, which I will discuss in this subsection.

At a given chemical potential $\mu$ in 2+1 flavor QCD with the fixed strange quark mass $m_s$,
the second order phase transition appears at $m_q= m_c(\mu)$ where $m_q$ is the light (u,d) quark mass. Furthermore the phase transition is of first order at $m_q < m_c(\mu)$ while it becomes a cross-over at  $m_q > m_c(\mu)$. The statement that the phase transition in 2+1 flavor QCD at $\mu=0$  is a cross-over at physical $m_q$ and $m_s$ correspond to the fact that
$m_q^{\rm phys} > m_c(0)$ at $m_s= m_s^{\rm phys}$ where $m_q^{\rm phy}$ and $m_s^{\rm phys}$ are physical light and strange quark masses, respectively.
If $m_c(\mu)$ becomes larger than $m_c(0)$ at larger $\mu$, it is expected that $m_c(\mu)$ at $m_s^{\rm phys}$ becomes equal to $m_q^{\rm phys}$ at some $\mu=\mu_c$;
$m_q^{\rm phys} = m_c(\mu_c)$. This $\mu_c$ corresponds to the value of the chemical potential at the critical end point of QCD. If this scenario is correct, it is expected that $m_c(\mu)$  increases from $m_c(0)$, so that $ \displaystyle\frac{\partial^2 m_c(\mu)}{\partial \mu^2} > 0$ at $\mu=0$. (Note that
$\displaystyle \frac{\partial m_c(\mu)}{\partial \mu} = 0$ at $\mu=0$.)

The imaginary chemical potential method avoids the complex phase problem of  lattice QCD with non-zero $\mu$ by taking $\mu = i \mu_I$, and lattice QCD with real $\mu_I$ can be easily simulated.
If $m_c(\mu)$ can be analytically continued to $\mu= i\mu_I$, we have
\begin{eqnarray}
\frac{m_c(\mu=i\mu_I)}{m_c(0)} &=& 1 + \sum_{k=1} c_k \left(\frac{\mu}{\pi T}\right)^{2k}
= 1+ \sum_{k=1} c_k (-1)^k\left(\frac{\mu_I}{\pi T}\right)^{2k}.
\end{eqnarray}
Therefore $c_k$ can be extracted from $m_c(\mu_I)$.
The latest result\cite{imaginary}  in 3 flavor QCD with $m_q=m_s$ at $N_t=4$
yields
\begin{eqnarray}
\frac{m_c(\mu)}{m_c(0)} &=& 1 -3.3(3) \left(\frac{\mu}{\pi T}\right)^{2} - 47(20)
\left(\frac{\mu}{\pi T}\right)^{4} +\cdots ,
\end{eqnarray}
which strongly indicates that the expected critical endpoint is absent at small $\mu/T$.
Of course it is necessary to check this unexpected behaviour by other methods for the definite conclusion.

\section{Potential between baryons}
\begin{figure}[bt]
\centering
\includegraphics[width=60mm,clip]{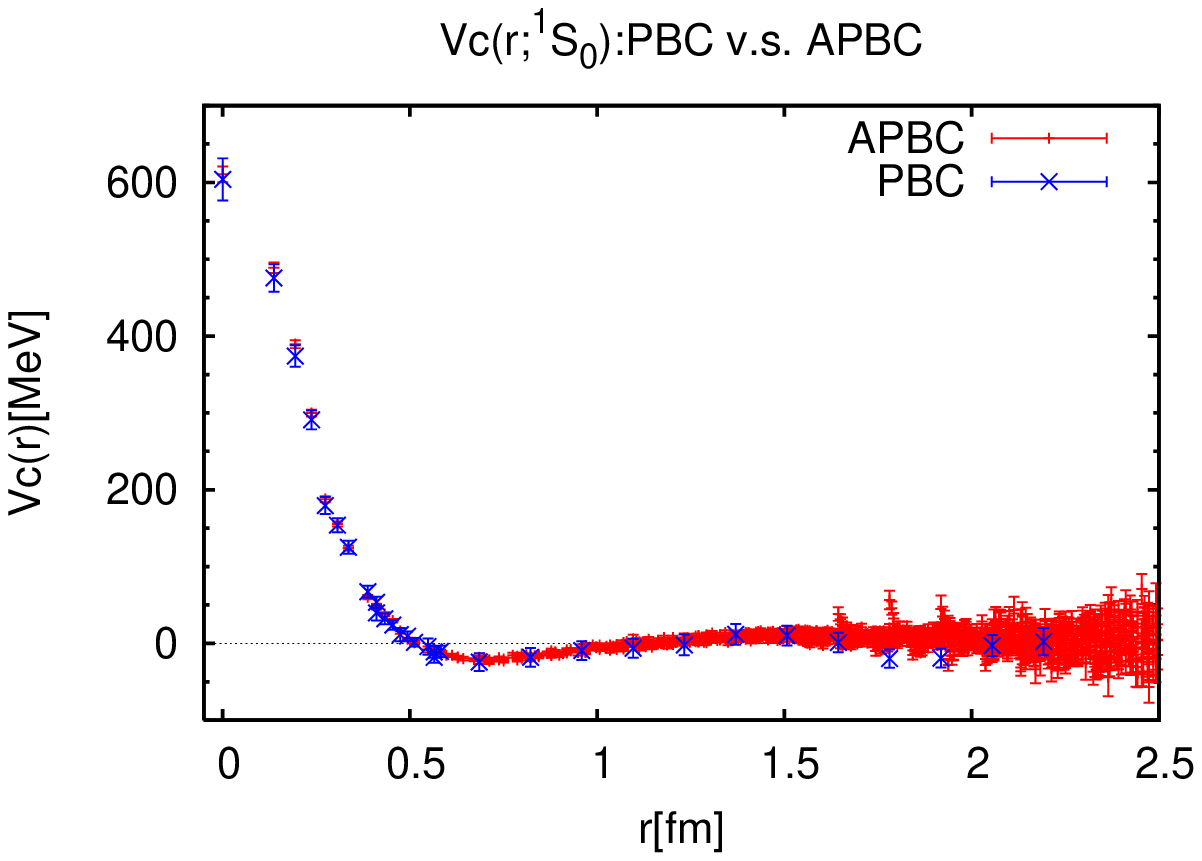}
\includegraphics[width=60mm,clip]{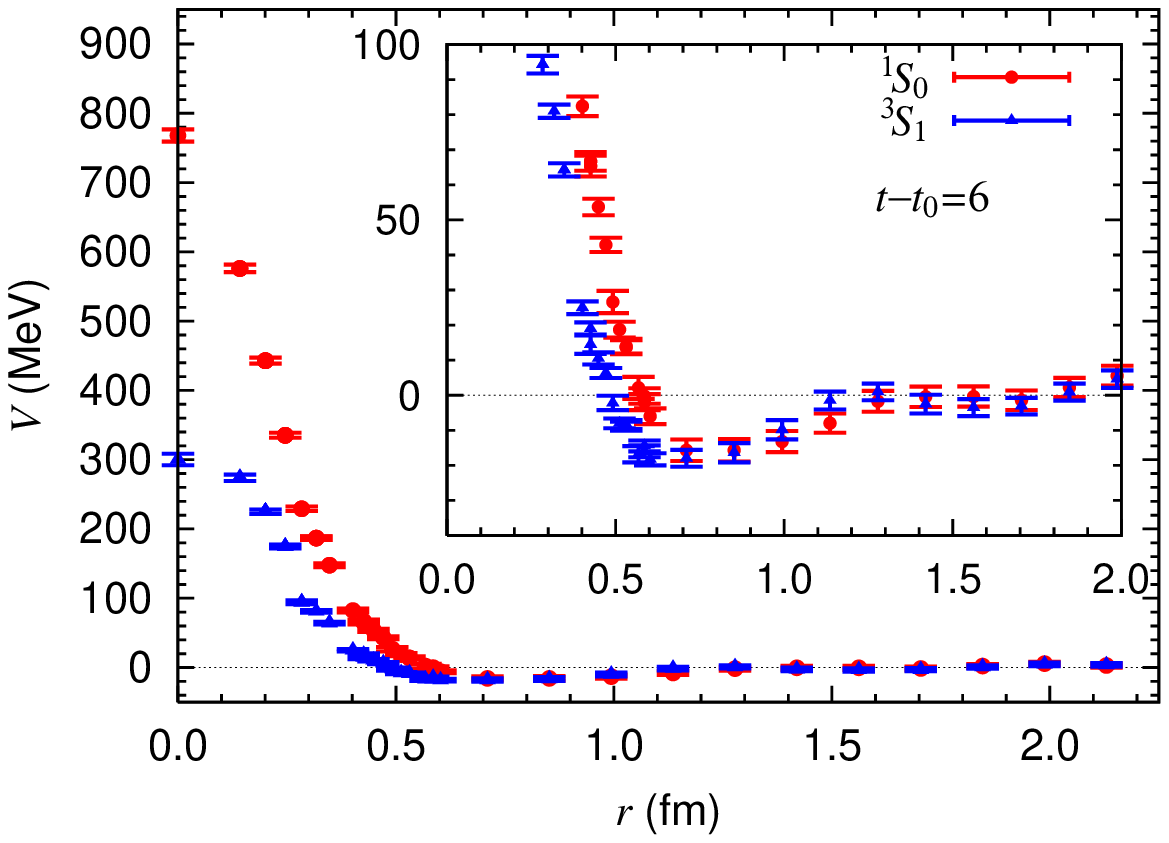}
\caption{ Left: The central potential for the $^1S_0$ 2N state with anti-periodic boundary condition
at $E\simeq 50$ MeV(red bars) and periodic boundary condition (blue crosses) at $E\simeq 0$ in quenched QCD. 
Right: The central potentials  for $^1S_0$ (red) and $^3S_1$ (blue) $I=1$ $N\Xi$ states in quenched QCD.}
\label{fig:pot}
\end{figure}

The force between nucleons (nuclear force) play a central role  in nuclear physics.
In a recent paper\cite{nforce}, which has received general recoginition\cite{nature},
the  potential between nucleons has been calculated from lattice QCD in the quenched approximation.  The result qualitatively reproduces all the features of phenomenological 
potentials that the force at medium to long range is attractive while  it becomes repulsive at short distance forming a characteristic repulsive core. 

The  potential between nucleons, $V({\bf x})$, is extracted from
the wave function $\varphi_E({\bf x}) $ through the Schr\"odinger equation as
\begin{eqnarray}
V({\bf x}) &=&\frac{ [E-H_0]\varphi_E ({\bf x})}{\varphi_E({\bf x})}, \quad H_0 = \frac{-{\bf \nabla}^2}{2\mu}, \quad
\varphi_E({\bf x}) = \langle 0 \vert N({\bf x}, 0) N({\bf 0}, 0) \vert 2N, E\rangle,
\end{eqnarray}
where $N({\bf x}, t)$ is a nucleon interpolating field, $\vert 2N, E\rangle$ is a 2 nucleon (2N) eigenstate in QCD with energy $E$ and $\mu$ is a reduced mass of 2N system ($\mu= m_N/2$).
In Ref.\cite{nforce}, $E\simeq 0$ has been used to extract the potential, but
the potential from this definition may depend on the energy $E$ of the 2N state.
In Fig.\ref{fig:pot} (Left), the potentials obtained at $E\simeq 0$(blue) and at $E\simeq 50$ MeV(red)
are compared in quenched QCD\cite{ABHIMNW}.  The potentials are almost identical between two energies, tough statistical fluctuations becomes sizable at large $r$ in the case of $E\simeq 50$ MeV,
partly due to the contamination from higher energy excited states.
It seems that the energy dependence of the potential is weak at this energy range.

An important applications of this method is to calculate the potential between nucleon and hyperon or
between two hyperons, which can not be extracted directly from scattering experiments.
In Fig.\ref{fig:pot} (Right). the potentials between $N$ and $\Xi$
in the $I=1$ channel for $^1S_0$ (red) and $^3S_1$(blue) states in quenched QCD are presented\cite{NIAH}.
While overall features are similar to those of nuclear potentials, 
stronger spin dependence of the potentials is observed in this case. 

\label{}
%
%
%

%
\end{document}